\documentclass[sort&compress]{elsarticle}
\pdfoutput=1

\newcommand{\lxor}{\oplus}
\pdfoutput=1
\usepackage{url}
\usepackage{times}
\usepackage{ifthen}
\newboolean{acmformat}
\setboolean{acmformat}{false}   
 \usepackage{setspace}

\usepackage{amssymb}
\usepackage{graphicx}
\graphicspath{{../viewsize-data/}{../ngramhashing-code/}{../../lemur/viewsize-data/}} 

\usepackage{pslatex}

\newcommand{\qxor}{\oplus}  

\makeatletter

\usepackage[normalem]{ulem}
 \usepackage{verbatim}

\usepackage{pslatex}
\usepackage{amsmath}
\usepackage{ntheorem}
\newcommand{\komment}[1]{}
\newtheorem{proposition}{Proposition}
\newtheorem{lemma}{Lemma}
\newtheorem{corollary}{Corollary}
\newtheorem{theorem}{Theorem}

\theoremheaderfont{\sc}\theorembodyfont{\upshape}
\theoremstyle{nonumberplain}
\theoremseparator{}
\theoremsymbol{\rule{1ex}{1ex}}
\newtheorem{proof}{Proof}

	\newcommand{\danielsafe}[1]{\textcolor{red}{DL:}\scriptsize\textbf{#1}\normalsize}
	\newcommand{\daniel}[1]{\marginpar{\begin{singlespace}\singlespace\textcolor{red}{DL:}\scriptsize\textbf{#1}\normalsize\end{singlespace}}}
	\newcommand{\owen}[1]{\marginpar{\begin{singlespace}\singlespace\textcolor{green}{OK:}\scriptsize\textbf{#1}\normalsize\end{singlespace}}}

\renewcommand{\daniel}[1]{}%
\renewcommand{\danielsafe}[1]{}%

\renewcommand{\owen}[1]{}%

\usepackage{color}
\usepackage{hyperref}

\newcommand{\apriori}{a priori}  

\newcommand{\notconference}[1]{#1}
\newcommand{\conference}[1]{}%

\usepackage{ifpdf}

\usepackage{algorithmic}
\usepackage{algorithm}

\begin{document}

\begin{frontmatter}

\title{Recursive n-gram hashing is pairwise independent, at best
}

%
%


 \author[UQAM]{Daniel Lemire\corref{cor1}} \ead{lemire@acm.org}
 \author[UNB]{Owen Kaser} \ead{o.kaser@computer.org}

 \address[UQAM]{\scriptsize LICEF, Universit\'e du Qu\'ebec \`a Montr\'eal (UQAM), 100 Sherbrooke West, Montreal, QC, H2X 3P2 Canada
}

 \address[UNB]{\scriptsize Dept.\ of CSAS, University of New Brunswick, 100 Tucker Park Road, Saint John, NB, Canada}
 \cortext[cor1]{Corresponding author. Tel.: 00+1+514 987-3000 ext. 2835; fax: 00+1+514 843-2160.}

\begin{abstract}
Many applications
use sequences of $n$~consecutive symbols ($n$-grams).
Hashing these $n$-grams can be a performance bottleneck.
For more speed,
recursive hash families compute hash values
by updating previous values.
We prove that recursive hash families cannot be more than
pairwise independent.
While hashing by irreducible polynomials
is  pairwise
independent, our implementations either run in time $O(n)$ or
use an exponential amount of memory. As a more scalable alternative, we  make hashing by
cyclic polynomials
pairwise independent by ignoring $n-1$~bits.
Experimentally, we show that hashing by cyclic polynomials
is twice as fast as hashing by irreducible polynomials.
We also show that randomized Karp-Rabin hash families
are not pairwise independent.
\end{abstract}



\end{frontmatter}



\section{Introduction}

An $n$-gram is a consecutive sequence of $n$ symbols from an alphabet $\Sigma$. 
An $n$-gram hash function $h$ maps $n$-grams to numbers in $[0,2^L)$.
These functions have several applications from full-text matching~\cite{cohen1998haa,cohen1999mqr,306482}, pattern matching~\cite{tan2006fpm}, or language models~\cite{cardenal2002fast,zhang2002minimum,schwenk2007csl,li2007fam,talbot2007smoothed,talbot2007randomised,talbot2008randomized} to plagiarism detection~\cite{857699}.

To prove that a hashing algorithm must work well, we typically need
hash values to satisfy some statistical property.
 Indeed, a hash function
that maps all $n$-grams to a single integer would not be useful.
Yet, a single hash function is deterministic: it maps an
$n$-gram to a single hash value.
Thus, we may be able to choose the input data so that the hash values
are biased.
Therefore, we randomly pick a function
from a family $\mathcal{H}$ of
functions~\cite{carter1979uch}.

Such
a family
$\mathcal{H}$ is \textit{uniform} (over $L$-bits) if all hash values
are equiprobable. That is,  considering $h$ selected
uniformly at random from $\mathcal{H}$, we have $P(h(x)=y)=1/2^L$ for all $n$-grams $x$ and all hash values $y$.
This condition is weak; the family of constant functions ($h(x)=c$) is uniform%
\footnote{%
\owen{do we need the first sentence here?}
\daniel{A reviewer or a reader could object that our definition of uniformity is
restrictive. We actually look for a particular type of uniformity. Feel free to
rephrase, but I think we need some kind of reference to flajolet, say.}
\owen{I just made it a footnote, because it is still sort of stuck
out and messed up the flow - seemed to be a choppy side observation.}
We omit families uniform over an arbitrary interval $[0,b)$---not of the form $[0,2^L)$.
Indeed, several applications~\cite{flajolet1985pca,Gibbons2001}
require uniformity over $L$-bits.
}.

Intuitively, we would want that if an adversary knows the hash
value of one $n$-gram, it cannot deduce anything about
the hash value of another $n$-gram. For example,
with the family of constant functions, once we know one
hash value, we know them all.
The family $\mathcal{H}$ is \textit{pairwise independent} 
 if the hash value
of $n$-gram $x_1$ is independent from the hash value of
any other $n$-gram $x_2$. That is,
we have
$P(h(x_1)=y \land h(x_2)=z)= P(h(x_1)=y) P(h(x_2)=z)=1/4^L$
for all distinct $n$-grams $x_1$, $x_2$, and all hash values $y$, $z$ with $x_1 \not = x_2$.
 Pairwise independence implies uniformity.
We refer to a particular hash function $h \in \mathcal{H}$ as ``uniform'' or ``a pairwise independent
hash function'' when the family in question can be inferred from the
context. \daniel{Basically: warning reader, we are sloppy.}

Moreover, the idea of pairwise independence can be generalized:
a family of hash functions $\mathcal{H}$ is \emph{$k$-wise independent}
if given distinct $x_1,\ldots,x_k$ and given $h$ selected
uniformly at random from $\mathcal{H}$,  then
$P(h(x_1)=y_1 \land \cdots \land h(x_k)=y_k)=1/2^{kL}$.  Note that $k$-wise
independence implies $k-1$-wise independence and uniformity.
(Fully) independent families are $k$-wise independent for arbitrarily large
$k$. For applications, non-independent families may fare as well as fully independent families
if the entropy of the data source is sufficiently high~\cite{1347164}.

A hash function $h$ is \emph{recursive}~\cite{cohenhash}---or rolling
\cite{winnowing}---if there is
a function $F$ computing the hash value of the
$n$-gram $x_2\ldots x_{n+1}$ from the hash value of
the preceding $n$-gram ($x_1 \ldots x_n$) and the
values of $x_1$ and $x_{n+1}$. That is, we have
\[h(x_2,\ldots,x_{n+1})=F(h(x_1,\ldots,x_n),x_1,x_{n+1}).\]
Ideally, we could compute function $F$ in time $O(L)$ and
not, for example, in time $O(Ln)$. 

The main contributions of this paper are:
\begin{itemize}
\item a proof that recursive hashing is no more than pairwise independent (\S~\ref{section:not3wise});
\item a proof that randomized Karp-Rabin can be uniform but never pairwise independent (\S~\ref{sec:id37});
\item a proof that hashing by irreducible polynomials is pairwise independent (\S~\ref{sec:general});
\item a proof that hashing by cyclic polynomials is not
even uniform (\S~\ref{sec:cyclic});
\item a proof that hashing by cyclic polynomials is
pairwise independent---after ignoring $n-1$~consecutive bits (\S~\ref{cyclicalmost}).
\end{itemize}
We conclude with an experimental section where we show that hashing
by cyclic polynomials is faster than hashing by irreducible polynomials.
Table~\ref{sec:summaryofpaper} summarizes the algorithms presented.

\begin{table}
\caption{\label{sec:summaryofpaper}A summary of the hashing function presented and their properties. For \textsc{General} and \textsc{Cyclic}, we require $L\geq n$. To make \textsc{Cyclic} pairwise independent, we need to discard some bits---the resulting scheme is not formally recursive.
Randomized Karp-Rabin is uniform under some conditions. }
\begin{tabular}{cccc}
\hline
name & cost per $n$-gram & independence & memory use\\ \hline
non-recursive 3-wise (\S~\ref{sec:RecursiveHashing})              & $O(Ln)$ & 3-wise & $O(n L |\Sigma|)$\\
Randomized Karp-Rabin (\S~\ref{sec:id37})  & $O(L \log L 2^{O(\log^*L)})$ & uniform & $O(L |\Sigma|)$\\
\textsc{General} (\S~\ref{sec:general})  & $O(Ln)$ & pairwise & $O(L |\Sigma|)$\\
RAM-Buffered \textsc{General} (\S~\ref{sec:rambuffered})  & $O(L)$ & pairwise & $O(L |\Sigma| + L 2^n)$\\
\textsc{Cyclic} (\S~\ref{sec:cyclic}) & $O(L+n)$  & pairwise (\S~\ref{cyclicalmost}) & $O( (L+n) |\Sigma|)$\\
\hline \end{tabular}
\end{table}

\section{Trailing-zero independence}

Some randomized algorithms~\cite{flajolet1985pca,Gibbons2001} merely require
that the number of trailing zeroes be independent.
For example, to estimate the number of distinct $n$-grams in a large document
without enumerating them, we merely have to compute  maximal numbers of
leading zeroes $k$ among hash values~\cite{durand2003lcl}. Na\"ively, we may
estimate that if a hash value with $k$~leading zeroes is found, we
have $\approx 2^k$~distinct $n$-grams.
Such estimates might be useful because the number of distinct $n$-grams grows
large with $n$:
Shakespeare's First Folio~\cite{Gutenberg} has over 3~million \textbf{distinct} 15-grams.

Formally, let $\textrm{zeros}(x)$ return the number of
trailing zeros (0,1,\ldots,$L$) of $x$, where $\textrm{zeros}(0) = L$.
We
say $h$ is \emph{$k$-wise trailing-zero independent} if $P(
\textrm{zeros}(h(x_1)) \geq j_1 \wedge \textrm{zeros}(h(x_2)) \geq j_2
\wedge \ldots \wedge \textrm{zeros}(h(x_k)) \geq j_k) =
2^{-j_1-j_2-\cdots-j_k}$, for $j_i = 0, 1, \ldots, L$.

If $h$
is $k$-wise independent, it is $k$-wise trailing-zero independent.
The converse is not true. If $h$ is a $k$-wise independent function,
consider $g \circ h$ where $g$ makes zero all bits before the
rightmost 1 (e.g., $g(0101100) = 0000100$). Hash $g \circ h$ is
$k$-wise trailing-zero independent but not even uniform (consider that
$P(g=0001)=8 P(g=1000)$).

\section{Recursive hash functions are no more than pairwise independent}
\label{section:not3wise}
Not only are recursive hash functions limited to pairwise
independence: they cannot be
3-wise \emph{trailing-zero} independent.

\daniel{Some reviewer wrote about the next proposition: Interesting observation. (\ldots) this
should lead us to \textbf{not} consider hash functions that are recursive (over hash
values).
}

\begin{proposition}\label{threewiseprop}
There is no 3-wise trailing-zero independent hashing function that is recursive.
\end{proposition}
\begin{proof}
\newcommand{\abfont}[1]{\texttt{#1}}
Consider the \owen{was: string of symbols, making
it clear what the exponent means} ($n+2$)-gram $\texttt{a}^n\texttt{bb}$.
Suppose $h$ is
recursive and
$3$-wise trailing-zero independent, then
\begin{eqnarray*}
\lefteqn{P \left (\textrm{zeros}(h(\abfont{a},\ldots,\abfont{a}))\geq L\bigwedge \right.}\\
& &\left.
    \textrm{zeros}(h(\abfont{a},\ldots, \abfont{a},\abfont{b})) \geq L \bigwedge
    \textrm{zeros}(h(\abfont{a},\ldots, \abfont{a},\abfont{b},\abfont{b})) \geq L  \right )\\
&= & P\left ( h(\abfont{a},\ldots,\abfont{a})=0 \bigwedge F(0,\abfont{a},\abfont{b})=0 \bigwedge F(0,\abfont{a},\abfont{b})=0 \right)\\
&= & P\left (h(\abfont{a},\ldots,\abfont{a})=0 \bigwedge F(0,\abfont{a},\abfont{b})=0\right )\\
& =& P\left (\textrm{zeros}(h(\abfont{a},\ldots,\abfont{a}))\geq L \bigwedge
    \textrm{zeros}(h(\abfont{a},\ldots, \abfont{a},\abfont{b})) \geq L\right )\\
& = &2^{-2L} \mbox{\ by trailing-zero pairwise independence}\\
& \not =& 2^{-3L} \mbox{\ as required by trailing-zero 3-wise independence.}
\end{eqnarray*}
Hence, we have a contradiction and no such $h$ exists.
\end{proof}

\section{A non-recursive 3-wise independent hash function}
\label{sec:RecursiveHashing}

A trivial way to generate an independent hash is to  assign a
random integer in $[0,2^L)$ to each new value $x$. Unfortunately, this
requires as much processing and storage as a complete indexing of all
values.


\daniel{One reviewer wrote:``
- P13, 26-28: This hash function can be shown to be 3-wise independent, but is
not 4-wise independent. Consider the 4 inputs (0,0,0,\ldots,0), (0,1,0,\ldots,0),
(1,0,0,\ldots,0), (1,1,0,\ldots,0). The bitwise XOR of their hash values is always 0,
so they are not independent.''}

 However, in a multidimensional setting this approach can be
put to good use. Suppose that we have tuples in
$K_1 \times K_2
\times \cdots \times K_n$ such that $\vert K_i \vert$ is small for
all $i$. We can  construct 
  independent hash functions
$h_i : K_i \rightarrow [0,2^L)$ for all $i$ and combine them.
The hash function
$h(x_1,x_2,\ldots,x_n)=h_1(x_1)\qxor h_2(x_2) \qxor \cdots \qxor
h_n(x_n)$ is then 3-wise 
 independent ($\qxor$ is the ``exclusive or'' function, XOR).
\newcommand{\xortextname}{XOR}
In time $O(\sum_{i=1}^n \vert K_i \vert)$,
we can construct the hash function by generating
$\sum_{i=1}^n \vert K_i \vert$ random numbers  and storing them in a look-up table.
With constant-time look-up, hashing an $n$-gram thus takes
$O(L n)$ time. Algorithm~\ref{algo:nwise} is an application
of this idea to $n$-grams.

\begin{algorithm}
\begin{algorithmic}[1]
\REQUIRE $n$  $L$-bit hash functions $h_1, h_1, \ldots, h_n$ over $\Sigma$ from an independent hash family
\STATE $s\leftarrow$ empty FIFO structure
\FOR{each character $c$}
\STATE append $c$ to $s$ \danielsafe{for some reason we use $s_i$ and not $x_i$}
\IF{length($s$)$= n$}
\STATE \textbf{yield}
 $h_1(s_1) \qxor h_2(s_2) \qxor \ldots  \qxor h_n(s_n)$ \\
\COMMENT{{\footnotesize The yield statement returns the value,
 without terminating the algorithm.}}
\STATE remove oldest character from $s$
\ENDIF
\ENDFOR
\end{algorithmic}
\caption{\label{algo:nwise}The (non-recursive) 3-wise 
 independent family.
 \danielsafe{One reviewer wrote: `` Algorithm 2: The way I read this, the algorithm is operating on n-bit numbers,
which seems to defy the purpose, leading to running time that is (not much
smaller than) O(N n). Is it not possible to do the computation modulo $x^L$, using
L-bit integers? (Same thing for algorithm 3.) ''}}
\end{algorithm}

This new family is not 4-wise independent for $n>1$. Consider
the  $n$-grams \texttt{ac},\texttt{ad},
\texttt{bc}, \texttt{bd}. The \xortextname\  of their
four hash values is zero. However, the family is 3-wise independent.

\owen{In what follows (well, proof of part A), there are some statements (regarding independence) that
are not true (I think) for the trivial case when $n=1$.  Have we explicitly
ruled this stupid case out somewhere, or can we slip in a condition quietly?
There is an explicit example using 1-grams!}
\daniel{No. If you see such a fault, it is a bug. Please outline it. I added
a precision below to clear up any misunderstanding.}
\begin{proposition}The family of hash functions
$h(x)= h_1(x_1)\lxor h_2(x_2) \lxor \ldots \lxor h_n(x_n)$,
where the $L$-bit hash functions $h_1,\ldots, h_n$ are
taken from an independent hash family,
is 3-wise independent.
\end{proposition}
\begin{proof}
Consider any 3~distinct $n$-grams: $x^{(1)}=x_1^{(1)}\ldots x_n^{(1)}$,
$x^{(2)}=x_1^{(2)}\ldots x_n^{(2)}$, and
$x^{(3)}=x_1^{(3)}\ldots x_n^{(3)}$.
Because the $n$-grams are distinct, at least one 
of two possibilities holds:
\begin{description}
\item[Case A] For some $i\in \{1,\ldots,n\}$, the three values $x_i^{(1)}, x_i^{(2)}, x_i^{(3)}$ are distinct. Write $\chi_j=h_i(x_i^{(j)})$ for $j=1,2,3$. For example, consider the three 1-grams: \texttt{a},\texttt{b},\texttt{c}.
\item[Case B] (Up to a reordering of the three $n$-grams.) There are two values $i,j\in \{1,\ldots,n\}$ such that $x_i^{(1)}$ is distinct from the two identical values $x_i^{(2)}, x_i^{(3)}$, and such that $x_j^{(2)}$ is distinct from the two identical values  $x_i^{(1)}, x_i^{(3)}$. Write $\chi_1=h_i(x_i^{(1)})$, $\chi_2=h_j(x_j^{(2)})$, and $\chi_3=h_i(x_i^{(3)})$.  For example, consider the three 2-grams: \texttt{ad},\texttt{bc},\texttt{bd}.
\end{description}

Recall that the \xortextname\  operation is invertible:
$a \lxor b = c$ if and only if $a = b \lxor c$.

We prove 3-wise independence for cases A and B.

\paragraph{Case A}

Write $f^{(i)} = h(x^{(i)})\lxor \chi_i$ for $i=1,2,3$. We have that
the values $\chi_1, \chi_2, \chi_3$ are mutually independent, and
they are
independent from the values $f^{(1)}, f^{(2)},f^{(3)}$\footnote{The values
$f^{(1)}, f^{(2)},f^{(3)}$ are not necessarily mutually independent.}:
\begin{eqnarray*}
P\left (\bigwedge_{i=1}^3  \chi_i=y_i \land \bigwedge_{i=1}^3  f^{(i)}=y'_i \right  )=\prod_{i=1}^3 P(\chi_i=y_i)P\left (\bigwedge_{i=1}^3  f^{(i)}=y'_i\right )
\end{eqnarray*}
for all values $y_i, y'_i$.
 Hence, we have
\begin{eqnarray*}
\lefteqn{P\left (h(x^{(1)})= z^{(1)} \bigwedge
h(x^{(2)})= z^{(2)} \bigwedge
h(x^{(3)})= z^{(3)}\right )} \\
&=& P\left (\chi_1 = z^{(1)}\lxor f^{(1)}) \bigwedge \chi_2 = z^{(2)} \lxor  f^{(2)}
\bigwedge \chi_3 = z^{(3)} \lxor  f^{(3)}\right )\\
&=& \sum_{\eta,\eta',\eta''} P\left (\chi_1 = z^{(1)}\lxor \eta \bigwedge \chi_2 = z^{(2)} \lxor \eta'
\bigwedge \chi_3 = z^{(3)} \lxor \eta'' \right ) \times \\
& &  P(f^{(1)}=\eta \land f^{(2)}=\eta' \land f^{(3)}=\eta'' ) \\
& = &  \sum_{\eta,\eta',\eta''} \frac{1}{2^{3L}} P(f^{(1)}=\eta \land f^{(2)}=\eta' \land f^{(3)}=\eta'') \\ 
& = & \frac{1}{2^{3L}}.
\end{eqnarray*}
Thus, in this case, the hash values are 3-wise independent.

\paragraph{Case B}

\owen{I note that case B precludes 1-grams, which is good because
otherwise some of the case B statements are false.}

Write $f^{(1)} = h(x^{(1)})\lxor \chi_1$,
$f^{(2)} = h(x^{(2)})\lxor \chi_2\lxor \chi_3$,
$f^{(3)} = h(x^{(3)})\lxor  \chi_3$.
Again,
the values $\chi_1, \chi_2, \chi_3$ are mutually independent, and
independent from the values $f^{(1)}, f^{(2)},f^{(3)}$.
We have
\begin{eqnarray*}
\lefteqn{P\left (h(x^{(1)})= z^{(1)} \bigwedge
h(x^{(2)})= z^{(2)} \bigwedge
h(x^{(3)})= z^{(3)}\right )} \\
&=& P\left (\chi_1 = z^{(1)}\lxor f^{(1)}) \bigwedge \chi_2 \lxor \chi_3 = z^{(2)} \lxor  f^{(2)}
\bigwedge \chi_3 = z^{(3)} \lxor  f^{(3)}\right )\\
&=& P\left (\chi_1 = z^{(1)}\lxor f^{(1)}) \bigwedge \chi_2  = z^{(2)} \lxor  f^{(2)}\lxor z^{(3)} \lxor  f^{(3)}
\bigwedge \chi_3 = z^{(3)} \lxor  f^{(3)}\right )\\
&=& \sum_{\eta,\eta',\eta''} P\left (\chi_1 = z^{(1)}\lxor \eta \bigwedge \chi_2 = z^{(2)}\lxor z^{(3)} \lxor \eta' \lxor \eta''
\bigwedge \chi_3 = z^{(3)} \lxor \eta'' \right ) \times \\
& &  P(f^{(1)}=\eta \land f^{(2)}=\eta' \land f^{(3)}=\eta'' ) \\
& = &  \sum_{\eta,\eta',\eta''} \frac{1}{2^{3L}} P(f^{(1)}=\eta \land f^{(2)}=\eta' \land f^{(3)}=\eta'') \\ 
& = & \frac{1}{2^{3L}}.
\end{eqnarray*}
This concludes the proof.
\end{proof}

\owen{I had a bad feeling in the proof of case B, when the $z^{(2)}$ and $z^{(3)}$ appeared in the same term.
I was thinking about how this statement was supposed to be true for all choices of these guys
and one could consider choosing them the same (and thereby cancelling them).  I eventually convinced myself that
you were still okay, but I'm definitely out of my league in this stuff.}
\daniel{What is above should be straight-forward. It is maybe hard to process because it is so formal,
but the math. is elementary, I think.
Yes, you can choose $z^{(2)}=z^{(3)}$. Ok. Let us work out this case. If $z^{(2)}=z^{(3)}$, then
$\chi_2 = z^{(2)}\lxor z^{(3)} \lxor \eta' \lxor \eta''= \eta' \lxor \eta''$.
And $\chi_3 = z^{(3)} \lxor \eta'' $. Ok, so let us substitute this back:
$\chi_2 \lxor \chi_3 = \eta' \lxor \eta''\lxor z^{(3)} \lxor \eta''= \eta' z^{(3)}= \eta' z^{(2)}$.
If you followed all that (congratulations!) then you should be convinced that it is perfectly ok. So
no problem occurs (that I can see) if $z^{(2)}=z^{(3)}$.}

\section{Randomized Karp-Rabin is not independent}
\label{sec:id37}
\daniel{We used to have a notation conflict regarding the
variable $B$. Make sure it does not happen again.}
One of the most common recursive hash functions is
commonly associated with the
Karp-Rabin string-matching algorithm~\cite{karp1987erp}. Given an integer $B$, the hash
value over the sequence of integers $x_1, x_2,\ldots, x_n$ is
$\sum_{i=1}^n x_i B^{n-i}$.
A variation of the Karp-Rabin hash method is ``Hashing by
Power-of-2 Integer Division''~\cite{cohenhash}, where
$h(x_1,\ldots,x_n) = \sum_{i=1}^n x_i B^{n-i} \bmod{2^L}$.
\daniel{We used to have the formula backward: $\sum_{i=1}^n x_i B^{n-i} \bmod{2^L}$. }
In particular, the \texttt{hashcode} method
of the Java String class uses this approach, with $L=32$ and
$B=31$~\cite{j15doc:String}. 
A widely used textbook~\cite[p. 157]{weis:dsaaj} recommends
a similar Integer-Division hash function for strings
with $B=37$.

Since such Integer-Division hash functions are  recursive,
quickly computed, and widely used, it is interesting to seek a
randomized version of them. Assume that $h_1$ is a random hash function
over symbols uniform in $[0,2^L)$, then
define
$h(x_1,\ldots,x_n)=B^{n-1} h_1(x_1)+B^{n-2} h_1(x_2)+ \cdots +  h_1(x_n)\bmod{2^L}$
for some fixed integer $B$. We choose $B=37$ (calling the resulting
randomized hash ``ID37;'' see Algorithm~\ref{algo:id37}). Our
algorithm computes each hash value in time
O($M(L)$), where $M(L)$ is the cost of multiplying two $L$-bit
integers. (We precompute the value $B^n \bmod{2^L}$.)
In many practical cases, $L$ bits can fit into a single machine word
and the cost of multiplication can be considered constant.  In general,
$M(L)$ is in $O(L \log L 2^{O(\log^*L)})$~\cite{1250800}.

 \daniel{We can also do it in $\lceil \log B^n \rceil L $ time if that's smaller.
 That's O(n L). Well, if we start saying that O(n L) is good, then there is no much
 point in being recursive, is there? I think we have to assume O(n L) is a bad thing,
 so O(L log L) is better. Not always true, of course... but in the spirit of the paper.}
\daniel{ If $2^L$ is small, we can precompute $B^n z$ for all $2^L$ possible values of $z$---thus reducing the complexity
 to $O(L)$. However, that is a useless remark since going to  $~L \log L$  to $L$ does not justify
 the memory buffer unless $L$ is huge, in which case the memory buffer won't work.}

\begin{algorithm}
\begin{algorithmic}[1]
\REQUIRE an $L$-bit hash function $h_1$ over $\Sigma$ from an independent hash family
\STATE $B\leftarrow 37$
\STATE $s\leftarrow$ empty FIFO structure
\STATE $x\leftarrow 0$ ($L$-bit integer)
\STATE $z\leftarrow 0$ ($L$-bit integer)
\FOR{each character $c$}
\STATE append $c$ to $s$
\STATE $x\leftarrow B x - B^n z + h_1(c) \bmod{2^L}$
\IF{length($s$)$= n$}
\STATE \textbf{yield} $x$
\STATE remove oldest character $y$ from $s$
\STATE $z \leftarrow h_1(y)$
\ENDIF
\ENDFOR
\end{algorithmic}
\caption{\label{algo:id37}The recursive ID37  family (Randomized Karp-Rabin). \danielsafe{I am not sure we should stress ID37 so much anymore, especially since choosing $B$ odd is bad as per our own theory?}}
\end{algorithm}

The randomized Integer-Division functions mapping  $n$-grams to $[0,2^L)$ are
 not pairwise independent. However, for some values of $B$ and $n$, they are uniform. 

\begin{proposition}
Randomized Integer-Division 
\daniel{I'd probably call it Randomized Karp-Rabin, but ok.}
 hashing  is not uniform for $n$-grams, if $n$ is even and $B$ is odd.
Otherwise, it is uniform for $B$ even and any $n$, or $B$ odd and $n$ odd. However, there is no value of $B$ for which it is  pairwise independent when $n\geq 2$.
\daniel{a reviewer wrote:  `This is a nice (though not too surprising) observation that
deserves to be known. ''}
\end{proposition}

\begin{proof}
\conference{Omitted; see \cite{viewsizetechreport}.}
\notconference{%
For $B$ odd,  we see that $P(h(\texttt{a}^{2k}) = 0) > 2^{-L}$ since
$h(\texttt{a}^{2k}) = h_1(\texttt{a}) ( B^0(1+B) + B^2(1+B) + \cdots + B^{2k-2}(1+B)) \bmod 2^L$
and since $(1+B)$ is even, we have
$P(h(\texttt{a}^{2k}) = 0) \geq P(h_1(x_1)=2^{L-1} \lor h_1(x_1)=0)= 1/2^{L-1}$.  Hence, for $B$ odd and $n$ even, we do not have uniformity.

Suppose that $B$ and $n$ are both odd.
Consider any string $x_1, x_2, \ldots, x_n$. We can find a character value $x_j$ which is repeated an odd number of times in the string. Let $I$ be the
set of indexes $i$ such that $x_i=x_j$.
We have that the equation $h(x_1, x_2, \ldots, x_n)=y$
is equivalent to
$(\sum_{i=1}^n B^{n-i} h_1(x_i))=y$. We
can rewrite it as
$(\sum_{i\in I} B^{n-i}) h_1(x_j)=y - (\sum_{i\not \in I} B^{n-i} h_1(x_i))$.
There is a unique solution $h_1(x_j)$ to this equation because
$(\sum_{i\in I} B^{n-i})$ is odd: the sum
of an odd number of odd integers is an odd integer. Hence, we have uniformity when $B$ and $n$ are odd.

Consider $B$ even. Consider any string $x_1, x_2, \ldots, x_n$. We are interested in the last character $x_n$. It might be repeated several times in the string. Let $I$ be the
set of indexes $i$ such that $x_i=x_n$.
We have that $h(x_1, x_2, \ldots, x_n)=y$
is equivalent to
$(\sum_{i=1}^n B^{n-i} h_1(x_i))=y$
or
$(\sum_{i\in I} B^{n-i}) h_1(x_n)=y - (\sum_{i\not \in I} B^{n-i} h_1(x_i))$.
We want to show that there is a unique solution $h_1(x_n)$ to this equation.
This follows because we have that $(\sum_{i\in I} B^{n-i})$ is an odd number because $B$ is even and $n\in I$. Hence, we have uniformity when $B$ is even.

To show it is not pairwise independent, first suppose that
$B$ is odd.  For any string $\beta$ of length $n-2$, consider
$n$-grams $w_1 = \beta \texttt{a} \texttt{a}$ and $w_2 = \beta \texttt{b} \texttt{b}$ for distinct
$\texttt{a}, \texttt{b} \in \Sigma$.
Then \owen{I changed 2 a's to b's.}\daniel{ok. good}
$P(h(w_1) = h(w_2)) = P(B^2 h(\beta) + B  h_1(\texttt{a})+h_1(\texttt{a})
= B^2 h(\beta) + B h_1(\texttt{b})+h_1(\texttt{b}) \bmod 2^L)
=P( (1+B) (h_1(\texttt{a})-h_1(\texttt{b})) \bmod 2^L = 0)\geq
P(h_1(\texttt{a})-h_1(\texttt{b}) = 0)+P(h_1(\texttt{a})-h_1(\texttt{b}) = 2^{L-1})$.
Because $h_1$ is independent, $P(h_1(\texttt{a})-h_1(\texttt{b})=0)=\sum_{c\in [0,2^L)} P(h_1(\texttt{a})=c)P(h_1(\texttt{b})=c)=\sum_{c\in [0,2^L)} 1/4^L=1/2^L$.
Moreover, $P(h_1(\texttt{a})-h_1(\texttt{b})= 2^{L-1})>0$. Thus,
we have that $P(h(w_1) = h(w_2))>1/2^L$ which contradicts pairwise independence.
Second, if $B$ is even, a similar argument shows
$P(h(w_3) = h(w_4)) > 1/2^L$, where
$w_3 = \beta \texttt{a} \texttt{a}$ and $w_4 = \beta \texttt{b} \texttt{a}$.
$P(h(\texttt{a},\texttt{a})=h(\texttt{b},\texttt{a}))=
P(B h_1(\texttt{a})+h_1(\texttt{a})=B h_1(\texttt{b})+h_1(\texttt{a}) \bmod 2^L)
=P(B(h_1(\texttt{a})-h_1(\texttt{b})) \bmod 2^L = 0)\geq
P(h_1(\texttt{a})-h_1(\texttt{b}) = 0)+P(h_1(\texttt{a})-h_1(\texttt{b}) = 2^{L-1})>1/2^L$.
Hence, as long as we consider strings of length $n>1$ and an alphabet $\Sigma$ containing at least two distinct characters, we can find two strings with a collision probability greater than $1/2^L$ whether $B$ is even or odd.
}
\end{proof}

A weaker condition than pairwise independence is 2-universality: a family is 2-universal if
$P(h(x_1)=h(x_2))\leq 1/2^L$~\cite{1347164}.
As a consequence of this proof, Randomized Integer-Division is not even 2-universal.

These results also hold for any Integer-Division hash where the modulo
is by an even number, not necessarily a power of 2.


\section{Generating hash families from  polynomials over Galois fields}
\label{sec:RecursiveHashingbyPolynomials}

A practical 
 form of hashing using the binary Galois
field GF(2) is called ``Recursive Hashing by Polynomials'' and has been
attributed to Kubina by Cohen~\cite{cohenhash}.
GF(2) contains only two values (1 and 0)  with the addition (and hence subtraction)
defined by \xortextname, $a + b = a \qxor b$ and the
multiplication by AND, $a \times b = a \wedge b$.
$\textrm{GF}(2)[x]$ is the
vector space of all polynomials with coefficients from GF(2).  Any
integer in binary form (e.g., $c=1101$) can thus be interpreted as an
element of $\textrm{GF}(2)[x]$ (e.g., $c=x^3+x^2+1$). If
$p(x)\in \textrm{GF}(2)[x]$,
then $\textrm{GF}(2)[x]/p(x)$ can be thought of as $\textrm{GF}(2)[x]$ modulo
$p(x)$.
As an example, if $p(x)=x^2$, then $\textrm{GF}(2)[x]/p(x)$ is the
set of all linear polynomials.  For instance, $x^3+ x^2+ x+1= x+1 \bmod{x^2}$ since,
in $\textrm{GF}(2)[x]$,
$(x+1) + x^2 (x+1) = x^3 + x^2 + x + 1$. \daniel{One reviewer pointed out that we use modulo to mean two things in this paper, and yet use the same notation. }

As a summary, we compute operations over $\textrm{GF}(2)[x]/p(x)$---where $p(x)$ is of degree $L$---as follows:
\begin{itemize}
\item the polynomial $\sum_{i=0}^{L-1} q_i x^i$ is represented as the $L$-bit integer $\sum_{i=0}^{L-1} q_i 2^i$;
\item subtraction or addition of two polynomials is the \xortextname\ of their $L$-bit integers;
\item multiplication of a polynomial $\sum_{i=0}^L q_i x^i$ by the monomial $x$ is represented either as $\sum_{i=0}^{L-1} q_i x^{i+1}$ if $q_{L-1}=0$
or as $p(x)+\sum_{i=0}^{L-1} q_i x^{i+1}$ otherwise. In other words, if the value of the last bit is 1, we merely apply a binary left shift, otherwise, we apply a binary left shift immediately followed by an \xortextname\ with the integer representing $p(x)$. In either case, we get an $L$-bit integer.
\end{itemize}
Hence, merely with the \xortextname\ operation, the binary left shift, and a way to evaluate the value of the last bit, we can compute all necessary operations over $\textrm{GF}(2)[x]/p(x)$ using integers.

Consider a hash function $h_1$ over characters taken from
some independent family. \daniel{Do we need $h_1$ to be really fully
independent or would n-wise or pairwise independent suffice? I think
that pairwise independence would not suffice.}
Interpreting $h_1$ hash values as polynomials in $\textrm{GF}(2)[x]/p(x)$,
and with the condition that $\textrm{degree}(p(x))\geq n$, we
define a hash function as  $h(a_1,a_2,\cdots,a_n)=h_1(a_1) x^{n-1} + h_1(a_2) x^{n-2}+
\cdots + h_1(a_n)$.  It \emph{is} recursive over the sequence
$h_1(a_i)$. The combined
hash can be computed \daniel{we used to say "computed in constant time with respect to $n$"} by reusing
previous hash values:
\[h(a_2,a_3,\ldots,a_{n+1})= x h(a_1,a_2,\ldots,a_n) - h_1(a_1) x^{n} + h_1(a_{n+1}).\]
Depending on the choice of the polynomial $p(x)$ we get different
hashing schemes, including \textsc{General} and \textsc{Cyclic}, which
are presented
in the next two sections.


\section{Recursive hashing by irreducible polynomials is pairwise independent}
\label{sec:general}

\begin{algorithm}

\begin{algorithmic}[1]
\REQUIRE an  $L$-bit hash function $h_1$ over $\Sigma$ from an independent hash family\danielsafe{not sure it must so strong!};
an irreducible polynomial $p$ of degree $L$ in $\textrm{GF}(2)[x]$

\STATE $s\leftarrow$ empty FIFO structure
\STATE $x\leftarrow 0$ ($L$-bit integer)
\STATE $z\leftarrow 0$ ($L$-bit integer)
\FOR{each character $c$}
\STATE append $c$ to $s$
\STATE $x \leftarrow \textrm{shift}(x)$
\STATE $z \leftarrow \textrm{shift}^n(z)$ \danielsafe{Here we have
a huge problem because this runs in $O(n)$ time, defeating the purpose
of recursive hashing in the first place!!!}
\STATE $x \leftarrow x \qxor z \qxor h_1(c)$
\IF{length($s$)$= n$}
\STATE \textbf{yield} $x$
\STATE remove oldest character $y$ from $s$
\STATE $z \leftarrow h_1(y)$
\ENDIF
\ENDFOR
\end{algorithmic}
\hrule
\begin{algorithmic}[1]
\STATE \textbf{function} shift
\STATE \textbf{input} $L$-bit integer $x$
\STATE shift $x$ left by 1~bit, storing result in an $L+1$-bit integer $x'$
\IF{leftmost bit of $x'$ is 1}
\STATE $x' \leftarrow x' \qxor p$
\ENDIF
\STATE \COMMENT{leftmost bit of $x'$ is thus always 0}
\STATE \textbf{return} rightmost $L$ bits of $x'$
\end{algorithmic}

\caption{\label{algo:general}The recursive \textsc{General}  family.}
\end{algorithm}

\begin{table}
\caption{\label{table:irred}Some irreducible polynomials over $\textrm{GF}(2)[x]$}
\centering
\begin{tabular}{cc}\hline
degree & polynomial  \\\hline
10     & $1+x^3+x^{10}$ \\
15     & $1+x+x^{15}$\\
20     & $1+x^3+x^{20}$\\
25     & $1+x^3+x^{25}$\\
30     & $1+x+x^4+x^6+x^{30}$\\ \hline
 \end{tabular}
 \end{table}
We can choose $p(x)$ to be an irreducible polynomial
of degree $L$ in
$\textrm{GF}(2)[x]$: an irreducible polynomial cannot be factored
into nontrivial polynomials (see Table~\ref{table:irred}).
 The
resulting hash is called 
\textsc{General} (see Algorithm~\ref{algo:general}). 
The main benefit of setting $p(x)$ to
be an irreducible polynomial is that $\textrm{GF}(2)[x]/p(x)$ is a field; in
particular, it is impossible that $p_1(x) p_2(x) = 0 \bmod
{p(x)}$ unless either $p_1(x)=0$ or $p_2(x)=0$. The field property
allows us to prove that the hash function is pairwise independent. 

\begin{lemma}\label{uniformlemma}
 \textsc{General} is pairwise independent.
 \daniel{One reviewer commented:``Lemma 2 and Theorem 5.1: I found these results interesting, since the hash
functions involve only rather simple bit operations. I wonder if some motivation
for these families could be found in efficient circuits for hashing? See, for
example, the paper "Static dictionaries on AC0 RAMs: query time Thetas(log
n/log log n) is necessary and sufficient", FOCS 96. ''}
\end{lemma}
\begin{proof}
If $p(x)$ is irreducible, then any non-zero $q(x)\in \textrm{GF}(2)[x]/p(x)$ has an inverse,
noted $q^{-1}(x)$ since
 $\textrm{GF}(2)[x]/p(x)$ is a field. Interpret hash values as polynomials in $\textrm{GF}(2)[x]/p(x)$.

Firstly, we prove that  \textsc{General} is uniform. In fact,
we show a stronger result: $P(q_1(x) h_1(a_1) + q_2(x) h_1(a_2)+\cdots+q_n(x)h_1(a_n)=y)= 1/2^L$
for any polynomials $q_i$ where at least one is different from zero.
The result follows by induction on the number of non-zero polynomials: it is clearly true
where there is a single non-zero polynomial $q_i(x)$,
since $q_i(x) h_1(a_i)=y \iff q_i^{-1}(x) q_i(x) h_1(a_i) = q_i^{-1}(x)y$.
Suppose it is true  up to $k-1$~non-zero
polynomials and consider a case where we have $k$~non-zero polynomials.
Assume without loss of generality that
$q_1(x)\neq 0$, we have
$P(q_1(x) h_1(a_1) + q_2(x) h_1(a_2)+\cdots+q_n(x)h_1(a_n)=y)=
P( h_1(a_1)  = q_1^{-1}(x)(y - q_2(x) h_1(a_2)-\cdots-q_n(x)h_1(a_n)))
=\sum_{y'} P( h_1(a_1)  = q_1^{-1}(x)(y - y'))P(q_2(x) h_1(a_2)+\cdots+q_n(x)h_1(a_n)=y')
= \sum_{y'} \frac{1}{2^L}\frac{1}{2^L}=\frac{1}{2^L}$ by the induction argument. Hence the
uniformity result is shown.

Consider two distinct sequences $a_1,a_2,\ldots,a_n$ and $a'_1,a'_2,\ldots,a'_n$.
Write $H_a= h(a_1,a_2,\ldots,a_n)$ and $H_{a'}=h(a'_1,a'_2,\ldots,a'_n)$.
 We have that
$P(H_a=y \land H_{a'}=y')
=P(H_a=y | H_{a'}=y')   P(H_{a'}=y')$.
Hence, to prove pairwise independence, it suffices to show that
 $P(H_a=y|H_{a'}=y')=1/2^L$.

Suppose that
$a_i= a'_j$ for some $i,j$; if not, the result follows since by the (full) independence
of the hashing function $h_1$, the values $H_a$
 and $H_{a'}$ are independent.
Write $q(x)= -(\sum_{k | a_k= a_i}  x^{n-k} ) (\sum_{k | a'_k= a'_j}  x^{n-k} )^{-1}$,
then
$H_a+q(x)H_{a'}$
is independent from $a_i= a'_j$ (and $h_1(a_i)=h_1(a'_j)$).

In $H_a+q(x)H_{a'}$,
only hashed values  $h_1(a_k)$ for $a_k \neq a_i$ and
$h_1(a'_k)$ for $a'_k \neq a'_j$ remain: label
them $h_1(b_1),\ldots, h_1(b_m)$. The result of the substitution
can be written $H_a+q(x)H_{a'} = \sum_k q_k(x) h_1(b_k)  $ where $q_k(x)$
are polynomials in $\textrm{GF}(2)[x]/p(x)$.
All $q_k(x)$ are zero if and only if $H_a+q(x)H_{a'}=0$
for all values of $h_1(a_1),\ldots, h_1(a_n)$ and $h_1(a'_1),\ldots, h_1(a'_n)$
(but notice that the value $h_1(a_i)=h_1(a'_j)$ is irrelevant);
in particular, it must be true when $h_1(a_k)=1$ and $h_1(a'_k)=1$ for all $k$, hence
$(x^n+\cdots+x+1)+q(x)(x^n\ldots+x+1)=0\Rightarrow q(x)=-1$. Thus,
all $q_k(x)$ are zero if and only if $H_a=H_{a'}$
for all values of $h_1(a_1),\ldots, h_1(a_n)$ and $h_1(a'_1),\ldots, h_1(a'_n)$ which only
happens if the sequences $a$ and $a'$ are identical.
Hence, not all $q_k(x)$ are zero.

Write $H_{y',a'}=(\sum_{k | a'_k= a'_j}  x^{n-k} )^{-1} (y'- \sum_{k | a'_k\neq a'_j}  x^{n-k} h_1(a'_k))$.
 On the one hand, the condition $H_{a'}=y'$ can be rewritten as
 $h_1(a'_j) = H_{y',a'}$. On the other hand,
$H_a+q(x)H_{a'}=y+q(x)y'$ is independent from $h_1(a'_j)=h_1(a_i)$.
Because $P(h_1(a'_j) = H_{y',a'})=1/2^L$ irrespective of $y'$ and
$h_1(a'_k)$ for $k\in \{k | a'_k\neq a'_j\}$, then
$P(h_1(a'_j) = H_{y',a'} | H_a+q(x)H_{a'}=y+q(x)y' )= P(h_1(a'_j) = H_{y',a'})$
which implies that $h_1(a'_j) = H_{y',a'}$
and $H_a+q(x)H_{a'}=y+q(x)y'$
are independent.
Hence, we have
\begin{eqnarray*}
\lefteqn{P( H_a=y | H_{a'}=y' )}&&\\
& = &P(  H_a+q(x)H_{a'}=y+q(x)y' | h_1(a'_j) = H_{y',a'})\\
& = &P(  H_a+q(x)H_{a'}=y+q(x)y')\\
& = &P(\sum_k q_k(x) h_1(b_k) = y + q(x) y'  )
\end{eqnarray*}
and by the earlier uniformity result, this last probability is equal to $1/2^L$.
This  concludes the proof.
\end{proof}

\section{Trading memory for speed: RAM-Buffered \textsc{General}}
\label{sec:rambuffered}
Unfortunately, \textsc{General}---as computed by Algorithm~\ref{algo:general}---requires $O(nL)$ time per $n$-gram.
Indeed, shifting a value $n$ times in
$\textrm{GF}(2)[x]/p(x)$ requires $O(nL)$~time.
However, if we are willing to trade memory usage for speed,
we can precompute these shifts. We call the resulting scheme
RAM-Buffered \textsc{General}.

\begin{lemma}\label{lemma:firsttradinglemma}
Pick any $p(x)$ in $\textrm{GF}(2)[x]$. The degree
of $p(x)$ is $L$.
Represent elements of  $\textrm{GF}(2)[x]/p(x)$
as polynomials of degree at most $L-1$.
Given any $h$ in
  $\textrm{GF}(2)[x]/p(x)$.
we can compute $x^n h$ in O($L$)~time
given an $O(L 2^n)$-bit memory buffer.
%
%
\end{lemma}
\begin{proof}
Write $h$ as $\sum_{i=0}^{L-1} q_i x^i$.
Divide $h$ into two parts,
$h^{(1)}=\sum_{i=0}^{L-n-1} q_i x^i$
and
$h^{(2)}=\sum_{i=L-n}^{L-1} q_i x^{i}$,
so that $h=h^{(1)}+ h^{(2)}$.
Then $x^n h  =
x^n h^{(1)} + x^n h^{(2)}$. The
first part, $x^n h^{(1)}$ is a polynomial
of degree at most $L-1$ since the degree of $h^{(1)}$ is at most
$L-1-n$. Hence, $x^n h^{(1)}$ as an $L$-bit value
is just $q_{L-n-1}q_{L-n-2}\ldots q_{0} 0 \ldots 0$.
which can be computed in time $O(L)$.
So, only the computation of
$x^n h^{(2)}$ is possibly more expensive than $O(L)$ time, but
$h^{(2)}$ has only $n$~terms as a polynomial (since the
first $L-n$~terms are always zero).
Hence, if we precompute $x^n h^{(2)}$ for all $2^n$~possible
values of  $h^{(2)}$, and store them in an array with $O(L)$~time
look-ups, we can compute $x^n h$
as an $L$-bit value in $O(L)$~time.
\end{proof}

When $n$ is large, this precomputation 
requires excessive space
and precomputation time.
Fortunately, we can trade back some speed for memory. 
Consider the proof of Lemma~\ref{lemma:firsttradinglemma}.
Instead of precomputing the shifts of all $2^n$~possible
values of  $h^{(2)}$ using an array of $2^n$~entries, we can further divide $h^{(2)}$ into $K$~parts.
For simplicity, assume that the integer $K$ divides $n$.
The $K$~parts $h^{(2,1)}, \ldots, h^{(2,K)}$
are made of the first $n/K$~bits,
the next  $n/K$~bits and so on. Because
$x^n h^{(2)}= \sum_{i=1}^K x^n h^{(2,i)}$, we can
shift $h^{(2)}$ by $n$ in $O(KL)$ operations using $K$~arrays
of  $2^{n/K}$~entries.
To summarize, we have a time complexity of $O(K L)$ per $n$-gram
using $O(L |\Sigma| + L K 2^{n/K})$~bits. We implemented the case $K=2$.

\section{Recursive hashing by cyclic polynomials is not even uniform}
\label{sec:cyclic}
Choosing $p(x)=x^{L}+1$ for $L\geq n$, for any polynomial
$q(x) = \sum_{i=0}^{L-1} q_i x^i$, we have
\[ x^i q(x) = x^i ( q_{L-1} x^{L-1}+  \cdots +q_1 x + q_0) = q_{L-i-1} x^{L-i-2}+ \cdots +  q_{L-i+1} x + q_{L-i}.\]
Thus, we have that multiplication by $x^i$ is a bitwise rotation, a cyclic left shift---which can be computed in $O(L)$~time.
The resulting hash (see Algorithm~\ref{algo:cyclic}) is called
\textsc{Cyclic}. 
It requires only $O(L)$~time per hash value.
\emph{Empirically}, Cohen showed that \textsc{Cyclic}
 is uniform~\cite{cohenhash}. In contrast, we show
 that it is not formally uniform:

\begin{algorithm}
\begin{algorithmic}[1]
\REQUIRE an  $L$-bit hash function $h_1$ over $\Sigma$ from an independent hash family\danielsafe{not sure it must so strong!}
\STATE $s\leftarrow$ empty FIFO structure
\STATE $x\leftarrow 0$ ($L$-bit integer)
\STATE $z\leftarrow 0$ ($L$-bit integer)
\FOR{each character $c$}
\STATE append $c$ to $s$
\STATE rotate $x$ left by 1~bit
\STATE rotate $z$ left by n~bits
\STATE $x \leftarrow x \qxor z \qxor h_1(c)$
\IF{length($s$)$= n$}
\STATE \textbf{yield} $x$
\STATE remove oldest character $y$ from $s$
\STATE $z \leftarrow h_1(y)$
\ENDIF
\ENDFOR
\end{algorithmic}
\caption{\label{algo:cyclic}The recursive \textsc{Cyclic}  family.}
\end{algorithm}

\begin{lemma}\label{not-uniformlemma}
 \textsc{Cyclic} is not uniform for $n$ even and never 2-universal, and thus never pairwise independent.
\end{lemma}
\begin{proof}

If $n$ is even, use the fact that $x^{n-1}+\cdots+x+1$ is divisible by $x+1$ to write
$x^{n-1}+\cdots+x+1=(x+1)r(x)$ for some polynomial $r(x)$. Clearly,
$r(x) (x+1)(x^{L-1}+x^{L-2}+\cdots+x+1)=0  \bmod{x^L+1}$ for any $r(x)$ and so
$P(h(a_1,a_1,\ldots, a_1)=0)=P((x^{n-1}+\cdots+x+1)h_1(a_1)=0)=
P((x+1) r(x) h_1(a_1)=0)
\geq P(h_1(a_1)=0 \lor h_1(a_1)=x^{L-1}+x^{L-2}+\cdots+x+1)= 1/2^{L-1}$.
Therefore,  \textsc{Cyclic} is not uniform for $n$ even.

To show \textsc{Cyclic} is never pairwise independent, consider $n=3$ (for simplicity),
then $P(h(a_1,a_1,a_2)= h(a_1,a_2,a_1))=P((x+1)(h_1(a_1)+h_1(a_2))=0)\geq
P(h_1(a_1)+h_1(a_2)=0 \lor h_1(a_1)+h_1(a_2)=x^{L-1}+x^{L-2}+\cdots+x+1)=1/2^{L-1}$,
but 
2-universal
hash values are equal with probability $1/2^L$. The result is shown. \end{proof}

Of the four
recursive hashing functions investigated by Cohen~\cite{cohenhash}, \textsc{General} and
\textsc{Cyclic} were
superior both in terms of speed and uniformity,
though \textsc{Cyclic} had a small edge over  \textsc{General}.
For $n$ large, the benefits of these recursive hash functions
compared to the 3-wise 
 independent hash function presented earlier
 can be substantial: $n$~table look-ups 
 is much
more expensive than a single look-up followed by binary shifts.

\section{\textsc{Cyclic} is pairwise independent if you remove $n-1$~consecutive bits}
\label{cyclicalmost}
\daniel{One reviewer said: `` 
In Section 5.2, the notation can be improved. The symbol "mod" is used
to denote two different functions, the first being the usual modulo
remainder, and the second being the dropping of a certain number of
bits from a string. This is very confusing, and makes this section
very hard to read, especially since the two meanings are used very
close to each other in the text.
''}
\daniel{One reviewer wrote: `` There is a slight ambiguity in the way this is
formulated. It could be understood as a property of single hash values (that any
two of its bits are independent) rather than as a property of any pair of hash
values.
''}
Because Cohen found empirically that \textsc{Cyclic} had good
uniformity~\cite{cohenhash}, it is reasonable to expect \textsc{Cyclic} to be
\textit{almost} uniform and maybe even \textit{almost} pairwise independent.
To illustrate this intuition, consider Table~\ref{tab:aa}  which shows
that while $h(\texttt{a},\texttt{a})$ is not uniform ($h(\texttt{a},\texttt{a})=001$ is impossible),
$h(\texttt{a},\texttt{a})$ minus any bit
is indeed uniformly distributed. We will prove that this result
holds in general. \daniel{The catch is that the resulting hash function
is no longer formally recursive. That is, once your start
"hiding" bits, you no longer follow Cohen's definition of
recursivity, and indeed, you pay a price since our complexity
becomes $O(n+L)$, instead of $O(L)$.}

\ifthenelse{\boolean{acmformat}}{%
\begin{acmtable}{0.87\columnwidth}
\centering\begin{tabular}{ccccc}
$h_1(\texttt{a})$ & $h(\texttt{a},\texttt{a})$ & $h(\texttt{a},\texttt{a})$ (first two bits) &
$h(\texttt{a},\texttt{a})$ (last two bits) & $h(\texttt{a},\texttt{a})$ (first and last bit)\\ \hline
 000 & 000 & 00 & 00& 00\\
 100 & 110 & 11 & 10& 10\\
 010 & 011 & 01 & 11& 01\\
 110&  101&  10 & 01 & 11\\
 001 &  101&  10 & 01& 11\\
 101 &  011&  01 & 11& 01\\
 011 &  110&  11 & 10& 10\\
 111 &  000&  00 & 00& 00\\
\end{tabular}
\caption{\label{tab:aa} \textsc{Cyclic} hash  for various values
of $h_1(\texttt{a})$ ($h(\texttt{a},\texttt{a})=xh_1(\texttt{a})+h_1(\texttt{a}) \bmod{2^L+1}$)}
\end{acmtable}%
}{%
\begin{table}
\caption{\label{tab:aa} \textsc{Cyclic} hash  for various values
of $h_1(\texttt{a})$ ($h(\texttt{a},\texttt{a})=xh_1(\texttt{a})+h_1(\texttt{a}) \bmod{2^L+1}$)}
\centering\begin{tabular}{ccccc}
\hline $h_1(\texttt{a})$ & $h(\texttt{a},\texttt{a})$ & $h(\texttt{a},\texttt{a})$ & $h(\texttt{a},\texttt{a})$ & $h(\texttt{a},\texttt{a})$ \\
 &  & (first two bits) &  (last two bits) &  (first and last bit)\\
\hline
 000 & 000 & 00 & 00& 00\\
 100 & 110 & 11 & 10& 10\\
 010 & 011 & 01 & 11& 01\\
 110&  101&  10 & 01 & 11\\
 001 &  101&  10 & 01& 11\\
 101 &  011&  01 & 11& 01\\
 011 &  110&  11 & 10& 10\\
 111 &  000&  00 & 00& 00\\
\hline
\end{tabular}
\end{table}
}

The next lemma and the next theorem show that \textsc{Cyclic} is quasi-pairwise independent
in the sense that $L-n+1$~consecutive bits (e.g., the first or last
$L-n+1$ bits) are pairwise independent. In other words,
\textsc{Cyclic} is pairwise independent if we are willing to sacrifice $n-1$~bits.
(We say that $n$~bits are ``consecutive modulo $L$'' if the bits are located at
indexes $i\bmod{L}$ for $n$ consecutive values of $i$ such as  $i=k,k+1,\ldots,k+n-1$.)

\begin{lemma}\label{lemma:magic} If $q(x) \in \textrm{GF}(2)[x]/(x^L+1)$
(with $q(x) \neq 0$)
has degree $n<L$, then
\begin{itemize}
\item the equation $q(x)w  = y \bmod{x^L+1}$ modulo the first $n$ bits\footnote
{By ``equality modulo
$\langle$\textit{some specified set of bit positions}$\rangle$'',
we mean that the two quantities are bitwise identical, with exceptions
permitted only at the specified positions.  For our polynomials, ``equality
modulo the first $n$ bit positions'' implies the difference of the two
polynomials has degree at most $n-1$.}
has exactly $2^n$~solutions for all $y$;
\item more generally, the equation $q(x)w  = y \bmod{x^L+1}$ modulo any consecutive $n$~bits
(modulo $L$) has exactly $2^n$~solutions  for all $y$.
\end{itemize}
\end{lemma}
\begin{proof}Let $P$ be the set of polynomials of degree at most $L-n-1$.
Take any $p(x)\in P$,
then $q(x)p(x)$ has degree at most $L-n-1+n=L-1$ and thus if $q(x) \neq 0$ and $p(x)\neq 0$,
then $q(x)p(x)\neq 0 \bmod{x^L+1}$. Hence, for any distinct $p_1,p_2\in P$ we have
$q(x) p_1 \neq q(x) p_2 \bmod{x^L+1}$.

To prove the first item, we begin by showing
that there is always exactly one solution in $P$.
Consider that there are $2^{L-n}$ polynomials $p(x)$ in $P$, and
that all values $q(x)p(x)$ are distinct. Suppose there are $p_1,p_2 \in P$ such
that $q(x)p_1 = q(x) p_2 \bmod{x^L+1}$ modulo the first $n$~bits, then
$q(x)(p_1 - p_2 )$ is a polynomial of degree at most $n-1$ while $p_1-p_2$ is a polynomial
of degree at most $L-n-1$ and $q(x)$ is
a polynomial of degree $n$, thus $p_1-p_2=0$.
(If $p1-p2 \neq 0$ then $\textrm{degree}(q(x)(p1-p2) \bmod{x^L+1}) \geq \textrm{degree}(q(x))=n$,
a contradiction.)
Hence, all $p(x)$ in $P$ are mapped to distinct values  modulo the first $n$~bits,
and since there are $2^{L-n}$ such distinct values, the result is shown.

Any polynomial of degree $L-1$ can be decomposed
into the form $p(x)+x^{L-n}z(x)$ where $z(x)$ is a polynomial
of degree at most $n-1$ and $p(x) \in P$. By the preceding result, for
distinct $p_1,p_2 \in P$,  $q(x)( x^{L-n}z(x)+p_1 )$ and
$q(x)( x^{L-n}z(x)+p_2 )$ must be distinct modulo the first $n$~bits.
In other words, the equation $q(x)( x^{L-n} z(x) + p ) = y$
modulo the first $n$~bits has exactly one solution $p\in P$ for
any $z(x)$ and since there are $2^n$~polynomials $z(x)$ of degree at most $n-1$,
then $q(x)w=y$ (modulo the first $n$~bits) must have $2^n$~solutions.

To prove the second item,
choose $j$ and use the first item to find any $w$ solving
$q(x)w =yx^j \bmod{x^L+1}$ modulo the
first $n$ bits.
$j$. Then $w x^{L-j}$ is a solution to $q(x)w =y \bmod{x^L+1}$ modulo
the bits in positions $j,j+1,\ldots,j+n-1 \bmod{L}$.
\end{proof}

We have the following corollary to Lemma~\ref{lemma:magic}.
\begin{corollary} \label{corollary:magic}If $w$ is chosen uniformly at random in
$\textrm{GF}(2)[x]/(x^L+1)$, then
$P(q(x) w = y\bmod{n-1 \textrm{\ bits}}) = 1/2^{L-n+1}$ where the $n-1$~bits
are consecutive (modulo $L$).
\end{corollary}

\begin{theorem} \label{thm:cyclicalmostind}
Consider the $L$-bit \textsc{Cyclic} $n$-gram hash family.
Pick any $n-1$~consecutive bit locations, then remove
these bits from all hash values.
The resulting $L-n+1$-bit hash family is pairwise independent.
\daniel{I rewrote the theorem to be less ambiguous.}
\end{theorem}
\begin{proof}

We show
$P(q_1(x) h_1(a_1) + q_2(x) h_1(a_2)+\cdots+q_n(x)h_1(a_n)=y \bmod{n-1 \textrm{\  bits}})= 1/2^{L-n+1}$
for any polynomials $q_i$ where at least one
is different from zero. It is  true
when there is a single non-zero polynomial
$q_i(x)$ by Corollary~\ref{corollary:magic}.
Suppose it is true up to $k-1$~non-zero
polynomials and consider a case where we have $k$~non-zero polynomials.
Assume without loss of generality that
$q_1(x)\neq 0$, we have
$P(q_1(x) h_1(a_1) + q_2(x) h_1(a_2)+\cdots+q_n(x)h_1(a_n)=y\bmod{n-1 \textrm{\  bits}})=
P( q_1(x) h_1(a_1)  = y - q_2(x) h_1(a_2)-\cdots-q_n(x)h_1(a_n)\bmod{n-1 \textrm{\ bits}})
=\sum_{y'} P(  q_1(x) h_1(a_1)  = y - y'\bmod{n-1 \textrm{\ bits}})
P(q_2(x) h_1(a_2)+\cdots+q_n(x)h_1(a_n)=y'\bmod{n-1 \textrm{\ bits}})
= \sum_{y'} \frac{1}{2^{L-n+1}} \frac{1}{2^{L-n+1}}=1/2^{L-n+1}$ by the induction argument,
where the sum is over $2^{L-n+1}$ values of $y'$.
Hence the uniformity result is shown.

Consider two distinct sequences $a_1,a_2,\ldots,a_n$ and $a'_1,a'_2,\ldots,a'_n$.
Write $H_a= h(a_1,a_2,\ldots,a_n)$ and $H_{a'}=h(a'_1,a'_2,\ldots,a'_n)$.
To prove pairwise independence, it suffices to show that
 $P(H_a=y \bmod{n-1 \textrm{\ bits}}|H_{a'}=y'\bmod{n-1 \textrm{\ bits}})=1/2^{L-n+1}$.
Suppose that
$a_i= a'_j$ for some $i,j$; if not, the result follows by the (full) independence
of the hashing function $h_1$. Using Lemma~\ref{lemma:magic},
find $q(x)$ such that
$q(x) \sum_{k | a'_k= a'_j}  x^{n-k} = - \sum_{k | a_k= a_i}  x^{n-k} \bmod{n-1 \textrm{\ bits}}$,
then
$H_a+q(x)H_{a'}  \bmod{n-1 \textrm{\ bits}}$
is independent from $a_i= a'_j$ (and $h_1(a_i)=h_1(a'_j)$).

The hashed values
$h_1(a_k)$ for $a_k \neq a_i$ and
  $h_1(a'_k)$ for $a'_k \neq a'_j$ are now relabelled as
 $h_1(b_1),\ldots, h_1(b_m)$. Write
 $H_a+q(x)H_{a'} = \sum_k q_k(x) h_1(b_k) \bmod{n-1 \textrm{\ bits}} $ where $q_k(x)$
are polynomials in $\textrm{GF}(2)[x]/(x^L+1)$  
(not all $q_k(x)$ are zero).
As in the proof of Lemma~\ref{uniformlemma}, we have that
$H_{a'}=y' \bmod{n-1 \textrm{\ bits}}$ and
$H_a+q(x)H_{a'}=y+q(x)y' \bmod{n-1 \textrm{\ bits}}$ are independent\footnote{%
We use the shorthand notation $P(f(x,y)=c | x, y)= b$ to
    mean $P(f(x,y)=c | x= z_1, y=z_2)=b$ for all values of $z_1, z_2$.
    }:
$P(H_{a'}=y' \bmod{n-1 \textrm{\ bits}} | y', b_1, b_2, \ldots, b_m)=1/2^{L-n+1}$ by
Corollary~\ref{corollary:magic} since $H_{a'}=y$ can be written as
$r(x) h_1(a'_j)=y- \sum_k r_k(x) h_1(b_k)$ for some polynomials $r(x), r_1(x), \ldots, r_m(x)$.
Hence, we have
\begin{eqnarray*}
\lefteqn{P( H_a=y \bmod{n-1 \textrm{\ bits}}| H_{a'}=y' \bmod{n-1 \textrm{\ bits}})}&&\\
& = &P(  H_a+q(x)H_{a'}=y+q(x)y' \bmod{n-1 \textrm{\ bits}}| H_{a'}=y' \bmod{n-1 \textrm{\ bits}})\\
& = &P(  H_a+q(x)H_{a'}=y+q(x)y' \bmod{n-1 \textrm{\ bits}})\\
& = &P(\sum_k q_k(x) h_1(b_k) = y + q(x) y' \bmod{n-1 \textrm{\ bits}} )
\end{eqnarray*}
and
by the earlier uniformity result, this last probability is equal to $1/2^{L-n+1}$.
\end{proof}

\section{Experimental comparison}

Irrespective of $p(x)$, computing
hash values has complexity  $\Omega(L)$. For \textsc{General}
and \textsc{Cyclic}, we require $L\geq n$. Hence,
the computation of their hash values is in $\Omega(n)$.
For moderate values of $L$ and $n$, this analysis is pessimistic because
CPUs can process 32- or 64-bit words in one operation.

To assess their real-world performance,
the various hashing algorithms\footnote{\url{http://code.google.com/p/ngramhashing/}.}
were written in C++. We compiled them with the GNU
GCC~4.0.1 compiler on an Apple MacBook with two Intel Core 2 Duo
 processors (2.4\,GHz) and 4\,GiB of RAM\@.
The -O3 compiler flag was used since it provided slightly better
performance for all algorithms. All hash values are stored
using 32-bit integers, irrespective of the number of bits used.

All hashing functions generate 19-bit hash values, except
for \textsc{Cyclic} which generates 19+$n$-bit hash values.
We had \textsc{Cyclic} generate more bits to compensate for
the fact that it is only pairwise independent after removal
of $n-1$~consecutive bits.
For \textsc{General}, we used the polynomial
$p(x)=x^{19}+x^{5}+x^{2}+x+1$~\cite{COS}.
For Randomized Karp-Rabin, we used the ID37 family.
The character hash-values are stored in an array for fast look-up.

\begin{figure}
\centering
\includegraphics[width=.6\textwidth]{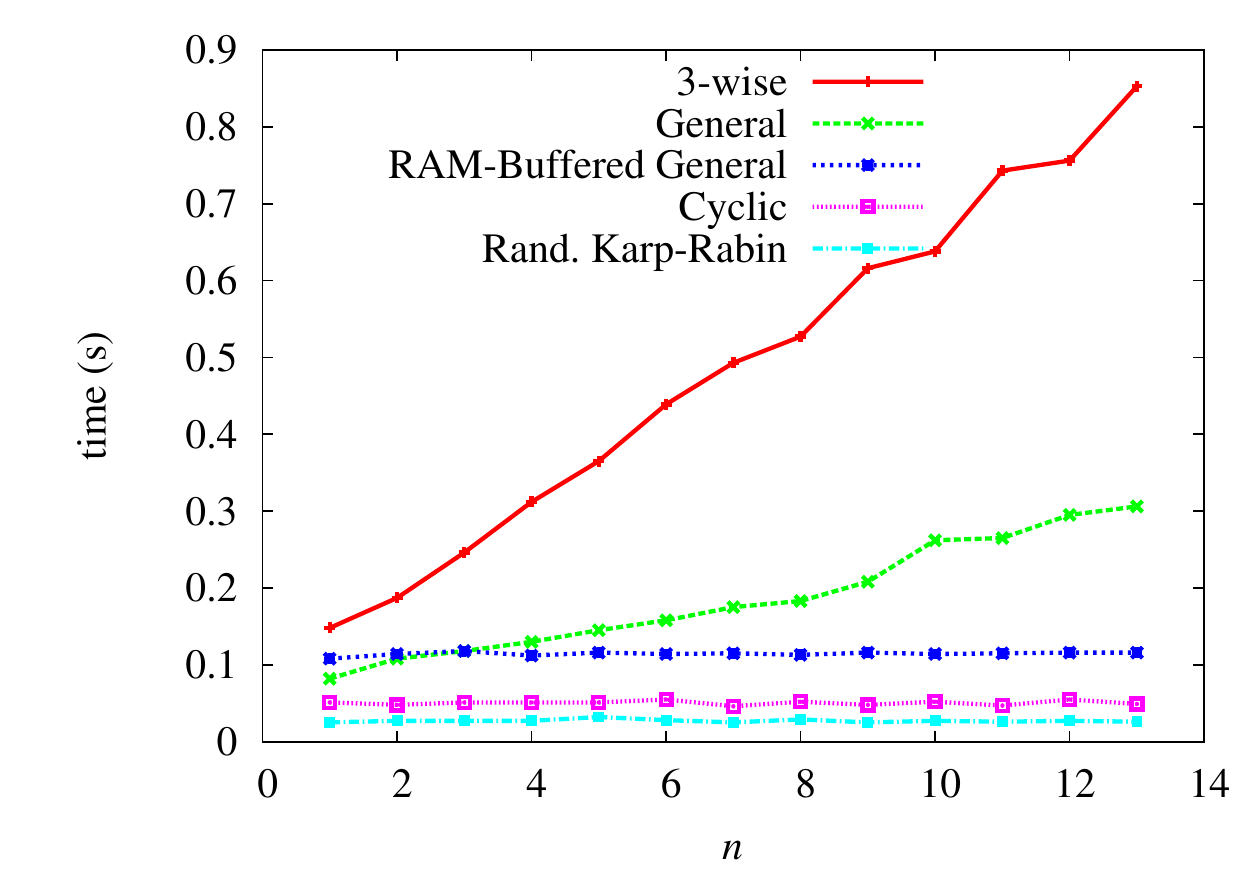}

\caption{\label{fig:timings}Wall-clock running time to hash all $n$-grams in the King James Bible}
\end{figure}

We report wall-clock time in Fig.~\ref{fig:timings}
for hashing the $n$-grams of the King James Bible~\cite{Gutenberg} which
contains 4.3~million ASCII characters.
\textsc{Cyclic} is twice as fast as \textsc{General}. 
As expected, the running time of the non-recursive hash function (3-wise)
grows linearly with $n$: for $n=5$, 3-wise is already  seven times slower than \textsc{Cyclic}.
Speed-wise, Randomized
Karp-Rabin (ID37) is the clear winner, being nearly twice as fast as \textsc{Cyclic}.
The performance of \textsc{Cyclic}
and ID37 is oblivious to $n$ in this test.

The RAM-Buffered \textsc{General} timings are---as expected---independent
of $n$, but they are twice as large as the \textsc{Cyclic} timings.
We do not show the modified version of RAM-Buffered \textsc{General}
that uses two precomputed arrays instead of a single one.
 It was
approximately 30\% slower than ordinary RAM-Buffered \textsc{General},
even up to $n=25$.
However, its RAM usage was 3~orders of magnitude smaller:
from 135\,MB down to 25\,kB.
Overall, we cannot recommend RAM-Buffered \textsc{General} or
its modification
considering that (1)~its memory usage
 grows
as $2^n$ and (2)~it is slower than \textsc{Cyclic}.

\section{Conclusion}

Considering speed and pairwise independence, we
recommend \textsc{Cyclic}---after discarding $n-1$~consecutive bits.
If we require only uniformity, Randomized Integer-Division is twice as fast.
\daniel{In particular, for $L$ small compared to $n$, we do not
know of a pairwise independent hash function computing hash values
in $O(L)$ time.}

\section*{Acknowledgments}

\owen{I think we need to be grateful to previous reviewers.
After all, they found the bug in the nwise stuff, commented on
3-wiseness.  Plus gave a lot of feedback that you used.}
\daniel{Ok. But I suggest we wait till the paper is accepted
otherwise the reviewers might find the following strange:
	"We would like to thank the numerous reviewers
who sank the several flawed papers we kept on submitting on this topic.
This paper is made entirely out of the good ideas these
reviewers provided. Alas, rules prohibit us from listing them
as the authors of this paper." }
This work is supported by NSERC grants 155967, 261437 and by  FQRNT
grant 112381.  The authors are grateful to the anonymous reviewers
for their significant contributions.

\bibliographystyle{elsarticle-num}
\bibliography{lemur}

\end{document}